\documentclass[11pt]{article}
\usepackage{amssymb,amsmath,amsfonts}
\usepackage{graphicx}
\usepackage{graphics}
\usepackage{eepic,epsfig}

\begin{document}

\title{Scalar and fermionic vacuum currents in de Sitter \\
spacetime with compact dimensions}
\author{ S. Bellucci$^{1}$\thanks{%
E-mail: bellucci@lnf.infn.it }, A. A. Saharian$^{2}$\thanks{%
E-mail: saharian@ysu.am }, H. A. Nersisyan$^{2}$ \vspace{0.3cm} \\
\textit{$^1$ INFN, Laboratori Nazionali di Frascati,}\\
\textit{Via Enrico Fermi 40, 00044 Frascati, Italy} \vspace{0.3cm}\\
\textit{$^2$ Department of Physics, Yerevan State University,}\\
\textit{1 Alex Manoogian Street, 0025 Yerevan, Armenia }}
\maketitle

\begin{abstract}
Vacuum expectation values (VEVs) of the current densities for charged scalar
and Dirac spinor fields are investigated in $(D+1)$-dimensional de Sitter
(dS) spacetime with toroidally compactified spatial dimensions. Along
compact dimensions we impose quasiperiodicity conditions with arbitrary
phases. In addition, the presence of a classical constant gauge field is
assumed. The VEVs of the charge density and of the components for the
current density along noncompact dimensions vanish. The gauge field leads to
Aharonov-Bohm-like oscillations of the components along compact dimensions
as functions of the magnetic flux. For small values of the comoving length
of a compact dimension, compared with the dS curvature scale, the current
density is related to the corresponding current in the Minkowski spacetime
by a conformal relation. For large values of the comoving length and for a
scalar field, depending on the mass of the field, two different regimes are
realized with the monotonic and oscillatory damping of the current density.
For a massive spinor field, the decay of the current density is always
oscillatory. In supersymmetric models on the background of Minkowski
spacetime with equal number of scalar and fermionic degrees of freedom and
with the same phases in the periodicity conditions, the total current
density vanishes due to the cancellation between the scalar and fermionic
parts. The background gravitational field modifies the current densities for
scalar and fermionic fields in different ways and for massive fields there
is no cancellation in the dS spacetime.
\end{abstract}

\bigskip

PACS numbers: 04.62.+v, 04.50.-h, 11.10.Kk, 04.20.Gz

\bigskip

\section{Introduction}

\label{sec:Introd}

The investigation of quantum effects in fixed gravitational backgrounds is
among the most interesting topics in quantum field theory (for reviews see
\cite{Birr82}). These effects may have important implications in black hole
physics and in cosmology. The presence of the gravitational field, in
general, reduces the number of symmetries and exact results can be obtained
for highly symmetric backgrounds only. A better understanding of physical
effects in these backgrounds could serve as a handle to deal with more
complicated geometries. In this context, the investigation of quantum field
theoretical effects in de Sitter (dS) spacetime is of special interest. In
inflationary scenario an approximately dS spacetime is employed to solve the
problem of initial conditions in standard cosmology \cite{Lind90}. During an
inflationary epoch in the early Universe the quantum fluctuations play a
crucial role in the generation of cosmic structures from inflation. More
recently observations of high-redshift Type Ia supernovae, galaxy clusters
and cosmic microwave background \cite{Ries07} indicate that at the present
epoch the Universe is accelerating and is well approximated by a model with
a positive cosmological constant as a dominant source for the expansion. If
the Universe would accelerate indefinitely, the standard cosmology would
lead to an asymptotic dS universe. Therefore, the investigation of physical
effects in dS spacetime is important for understanding both the early
Universe and its future. An interesting topic which has received increasing
attention is related to string-theoretical models of dS spacetime and
inflation. A number of constructions of metastable dS vacua were discussed
within the framework of string theory (see, for instance, \cite{Kach03}).

In recent years much attention has been paid to the possibility for the
universe to have non-trivial topology. In particular, a number of
fundamental physical theories are formulated in spacetimes with compact
extra dimensions. The idea of compactified dimensions has been extensively
used in supergravity and superstring theories. The models of a compact
universe with non-trivial topology may play an important role by providing
proper initial conditions for inflation \cite{Lind04}. The quantum creation
of the universe having toroidal spatial topology is discussed in \cite%
{Zeld84} and in Refs. \cite{Gonc85} within the framework of supergravity
theories. As it was argued in Refs. \cite{McIn04}, there is no reason to
believe that the version of dS spacetime which may emerge from string
theory, will necessarily be the most familiar version with symmetry group $%
O(1,4)$ and there are many different topological spaces which can accept the
dS metric locally. There are several reasons to expect that in string theory
the most natural topology for the universe is that of a flat compact
three-manifold.

The non-trivial topology of the background space can have important physical
implications in quantum field theory. The periodicity conditions imposed on
fields along compact dimensions give rise to the modification of the
spectrum for normal modes and, related to this, the expectation values of
physical observables are changed. A well known effect of this type,
demonstrating the connection between quantum phenomena and global properties
of spacetime, is the topological Casimir effect \cite{Most97,Duff86}. The
Casimir energy of bulk fields induces a non-trivial potential for the
compactification radius providing a stabilization mechanism for moduli
fields and thereby fixing the effective gauge couplings. The Casimir effect
has also been considered as a possible origin for the dark energy in both
Kaluza-Klein-type and braneworld models \cite{DarkEn}.

One-loop quantum effects for scalar and Dirac spinor fields induced by the
toroidal compactification of spatial dimensions in dS spacetime have been
recently investigated in Refs. \cite{Saha07,Bell08,Saha08,Beze09}. In these
papers the vacuum expectation values (VEVs) for the field squared and the
energy-momentum tensor are considered for untwisted and twisted fields.
These quantities are among the most important local characteristics of the
quantum vacuum and are closely related with the structure of spacetime
through the theory of gravitation. For charged fields another important
bilinear characteristic is the VEV of the current density. In addition to
describing the physical structure of the quantum field at a given point, the
current acts as the source in the Maxwell equations and plays an important
role in modeling a self-consistent dynamics involving the electromagnetic
field.

In the present paper, we consider the combined effects of topology and the
gravitational field on the VEVs of the current density for charged scalar
and fermionic fields in the background of dS spacetime with an arbitrary
number of toroidally compactified spatial dimensions. Along compact
dimensions we impose generic quasiperiodic boundary conditions with
arbitrary phases which, as special cases, include the periodicity conditions
for untwisted and twisted fields. In addition, the presence of a classical
constant gauge field will be assumed. As it will be seen, this leads to
Aharonov-Bohm-like effects on the VEV\ of the current density. The paper is
organized as follows. In the next section we describe the background
geometry and evaluate the Hadamard function for a complex scalar field
assuming that the field is prepared in the Bunch-Davies vacuum state. By
using this function, in section \ref{Sec:Sc}, we investigate the VEV of the
current density. The current density for the Dirac spinor field is discussed
in section \ref{Sec:Ferm}. The main results are summarized in section \ref%
{sec:Conc}.

\section{Hadamard function for a scalar field}

\label{Sec:Had}

In the present paper the background geometry is described by the $(D+1)$%
-dimensional dS line-element in planar coordinates:%
\begin{equation}
ds^{2}=dt^{2}-e^{2t/\alpha }\sum_{l=1}^{D}(dx^{l})^{2}.  \label{ds2}
\end{equation}%
In the case of trivial topology, the metric tensor corresponding to Eq. (\ref%
{ds2}) is the maximally symmetric vacuum solution of Einstein
equations with the cosmological constant $\Lambda =D(D-1)\alpha
^{-2}/2$. For the coordinates $\mathbf{x}_{p}=(x^{1},...,x^{p})$,
as usual, one has $-\infty
<x^{l}<\infty $, $l=1,\ldots ,p$, and here we assume that the coordinates $%
\mathbf{x}_{q}=(x^{p+1},...,x^{D})$, with $q=D-p$, are compactified to
circles with the lengths $L_{l}$: $0\leq x^{l}\leq L_{l}$, $l=p+1,\ldots ,D$%
. Hence, we consider the spatial topology $R^{p}\times (S^{1})^{q}$. The
compactification of a part of spatial coordinates reduces the symmetry of dS
spacetime (this is the case for Minkowski spacetime as well). But this does
not cause any problem from the cosmological point of view. From this
perspective the geometry under consideration is just a special case of the
larger class of spatially flat, Kaluza-Klein-type models which are relevant
to inflationary cosmology. Moreover, as it has been pointed out in Ref. \cite%
{Witt01} (see also \cite{McIn04,Klem04}), there is every reason to doubt
that quantum gravity effects are symmetric under the full dS group. In the
discussion below it will be convenient, in addition to the time coordinate $t
$, to use the conformal time $\tau =-\alpha e^{-t/\alpha }$ with $-\infty
<\tau <0$. For this coordinate the line element takes a conformally flat
form with the conformal factor $\left( \alpha /\tau \right) ^{2}$: $%
ds^{2}=\left( \alpha /\tau \right) ^{2}\eta _{\mu \nu }dx^{\mu }dx^{\nu }$,
where $\eta _{\mu \nu }$ is the Minkowskian metric tensor.

Though the spacetime under consideration is locally equivalent to the $(D+1)$%
-dimensional global dS spacetime with the spatial topology $S^{D}$, the
global properties of these manifolds are very different. The special case $%
p=0$, $q=D=3$, with all spatial dimensions been compactified and
referred as Spatially Toral de Sitter (STdS) spacetime, has been
previously discussed in the literature (see Refs.
\cite{Lind04}-\cite{McIn04}). It has been shown that the
modification of the topology of dS spacetime leads to important
physical consequences. Unlike to the familiar dS spacetime, STdS
is geodesically incomplete in the past \cite{Ande03}. This is
related to the fact that the planar coordinates do not cover the
full dS manifold. It is known that (see, for example,
\cite{Degu13} for a recent discussion) this does not cause
problems in the quantization of fields because $t=\mathrm{const}$
defines a Cauchy surface and the information from any part of dS
manifold enters the submanifold covered by planar coordinates as
an initial condition. As it has been pointed out in Refs.
\cite{Zeld84,Lind04}, important physical differences of closed dS
and STdS spacetimes arise in the discussion of quantum creation of
the universe. In a compact flat or open universe, unlike to the
case of a closed or an infinite open universe, for the theories
with the inflaton potential $V(\phi )\ll 1$ (in Planckian units)
the probability of inflation is not exponentially suppressed
\cite{Lind04}. For a closed dS universe, the effective potential
$V_{\mathrm{eff}}(a)$ ($a$ is the scale factor) in the
Wheeler-DeWitt equation contains a term proportional to $a^{2} $
which comes from the spatial curvature. This leads to the
existence of the barrier in $V_{\mathrm{eff}}(a)$. In STdS
spacetime such a barrier is absent and there is no need for
tunnelling through a classically forbidden region for the scale
factor. In this case the cosmological evolution can begin with an
arbitrarily small value of $a$ and there is no exponential
suppression of the probability for quantum creation of the
universe associated with tunnelling. In Ref. \cite{McIn04} it was
argued that the pre- and post-inflationary distortions of the STdS
spacetime do not give rise to non-perturbative instabilities in
string theory. In particular, it has been shown that there is no
danger of Seiberg-Witten instability \cite{Seib99} in toral string
cosmology as long as conventional matter dominates in the
post-inflationary evolution. This type of instability arises due
to the emission of "large branes". As it will be shown below, the
nontrivial topology of the patch of dS spacetime covered by planar
coordinates leads to another physical consequence, namely, to the
appearance of nonzero VEVs of the current density for charged
fields.

Firstly, we consider a complex scalar field $\varphi (x)$ with a curvature
coupling parameter $\xi $, in the presence of a classical abelian gauge
field $A_{\mu }$. The corresponding field equation has the form%
\begin{equation}
\left( D_{\mu }D^{\mu }+m^{2}+\xi R\right) \varphi (x)=0,  \label{FieldEq}
\end{equation}%
where $D_{\mu }=\nabla _{\mu }+ieA_{\mu }$ is the gauge-covariant derivative
and $e$ is the coupling between the scalar and gauge fields. For the scalar
curvature in (\ref{FieldEq}) one has $R=D(D+1)/\alpha ^{2}$. In the most
important special cases of minimally and conformally coupled scalars one has
$\xi =0$ and $\xi =(D-1)/(4D)$, respectively. Since the background space is
multiply-connected, in addition to the field equation one should specify the
periodicity conditions along compact dimensions. We will assume generic
quasiperiodic boundary condition%
\begin{equation}
\varphi (t,\mathbf{x}_{p},\mathbf{x}_{q}+L_{l}\mathbf{e}_{l})=e^{i\alpha
_{l}}\varphi (t,\mathbf{x}_{p},\mathbf{x}_{q}),  \label{BC_pq}
\end{equation}%
with constant phases $\alpha _{l}$ and with $\mathbf{e}_{l}$ being the unit
vector along the dimension $x^{l}$, $l=p+1,...,D$ (for physical effects of
phases in periodicity conditions for fields in multiply-connected spaces see
also \cite{Hoso83}). The results below will be periodic functions of $\alpha
_{l}$ with the period equal to $2\pi $. The special cases of untwisted and
twisted scalar fields (periodic and antiperiodic boundary conditions), most
frequently discussed in the literature, correspond to $\alpha _{l}=0$ and $%
\alpha _{l}=\pi $, respectively.

Here we will consider the simplest configuration of the gauge field with $%
A_{\mu }=\mathrm{const}$. In this case, the gauge field can be excluded from
the field equation by the gauge transformation
\begin{equation}
A_{\mu }^{\prime }=A_{\mu }+\partial _{\mu }\omega ,\;\varphi ^{\prime
}(x)=e^{-ie\omega }\varphi (x),\;\omega =-A_{\mu }x^{\mu }.  \label{Gauge}
\end{equation}%
In the new gauge one has $A_{\mu }^{\prime }=0$ and $D_{\mu }^{\prime
}=\nabla _{\mu }$. Now the quasiperiodicity condition takes the form%
\begin{equation}
\varphi ^{\prime }(t,\mathbf{x}_{p},\mathbf{x}_{q}+L_{l}\mathbf{e}_{l})=e^{i%
\tilde{\alpha}_{l}}\varphi ^{\prime }(t,\mathbf{x}_{p},\mathbf{x}_{q}),
\label{BC1}
\end{equation}%
where%
\begin{equation}
\tilde{\alpha}_{l}=\alpha _{l}+eA_{l}L_{l}.  \label{alf1}
\end{equation}%
As it is seen, the phases in the periodicity conditions and the value of the
gauge field are related to each other through a gauge transformation. In
what follows, the physical results will depend on the phases in the
periodicity conditions and on the gauge potential in the form of the
combination (\ref{alf1}). Note that if $\mathbf{A}$ is the spatial vector
with $D$ components corresponding to the spacetime vector $A_{\mu }$, then $%
\mathbf{A}_{l}=-A_{l} $. Although the corresponding field strength vanishes,
a constant gauge field shifts the phases in the periodicity conditions along
compact dimensions. As it will be seen below, this leads to the
Aharonov-Bohm-like effects on the current density for charged fields. Note
that in (\ref{alf1}) the shift due to the gauge field may be written in the
form $eA_{l}L_{l}=-e\mathbf{A}_{l}L_{l}=-2\pi \Phi _{l}/\Phi _{0}$, where $%
\Phi _{0}=2\pi /e$ is the flux quantum and $\Phi _{l}$ is the flux enclosed
by the circle corresponding to the $l$th compact dimension. In what follows
we will work in terms of the gauge transformed field $\varphi ^{\prime }(x)$
omitting the prime.

The physical quantity we are interested in is the current density for a
charged scalar field. The corresponding operator is given by the standard
expression:%
\begin{equation}
j_{\mu }(x)=ie[\varphi ^{+}(x)D_{\mu }\varphi (x)-(D_{\mu }\varphi
^{+}(x))\varphi (x)].  \label{jsc}
\end{equation}%
Note that in the gauge under consideration one has $D_{\mu }\varphi =\nabla
_{\mu }\varphi =\partial _{\mu }\varphi $. The VEV of the current density, $%
\left\langle j_{\mu }(x)\right\rangle =\left\langle 0\right\vert j_{\mu
}(x)\left\vert 0\right\rangle $, is expressed in terms of the Hadamard
function%
\begin{equation}
G^{(1)}(x,x^{\prime })=\left\langle 0\right\vert \varphi (x)\varphi
^{+}(x^{\prime })+\varphi ^{+}(x^{\prime })\varphi (x)\left\vert
0\right\rangle ,  \label{G1}
\end{equation}%
where $\left\vert 0\right\rangle $ stands for the vacuum state. The
corresponding formula reads:
\begin{equation}
\left\langle j_{\mu }(x)\right\rangle =\frac{i}{2}e\lim_{x^{\prime
}\rightarrow x}(\partial _{\mu }-\partial _{\mu }^{\prime
})G^{(1)}(x,x^{\prime }).  \label{jscH}
\end{equation}%
By expanding the field operator in terms of a complete set $\{\varphi
_{\sigma }^{(\pm )}(x)\}$ of solutions to the field equation, obeying the
periodicity conditions (\ref{BC1}), and using the standard commutation
relations for the annihilation and creation operators, the following formula
is obtained for the Hadamard function:%
\begin{equation}
G^{(1)}(x,x^{\prime })=\sum_{\sigma }\sum_{s=\pm }\varphi _{\sigma
}^{(s)}(x)\varphi _{\sigma }^{(s)\ast }(x^{\prime }).  \label{Had}
\end{equation}%
Here $\sum_{\sigma }$ includes the summation over the discrete components of
the collective index $\sigma $ and the integration over the continuous ones
(for the specification of the collective index in the geometry under
consideration see below).

Because of the plane symmetry of the problem, the dependence of the scalar
mode functions on spatial coordinates can be taken in the standard
exponential form, $e^{i\mathbf{k}\cdot \mathbf{x}}$, with $\mathbf{x=(x}_{p}%
\mathbf{,x}_{q}\mathbf{)}$ and $\mathbf{k=(k}_{p}\mathbf{,k}_{q}\mathbf{)}$.
From the field equation it can be seen that the general solution for the
time-dependent part of the mode functions is a linear combination of the
functions $\eta ^{D/2}H_{\nu }^{(1)}(k\eta )$ and $\eta ^{D/2}H_{\nu
}^{(2)}(k\eta )$, where $k=|\mathbf{k}|$,%
\begin{equation}
\eta =|\tau |=\alpha e^{-t/\alpha },  \label{eta}
\end{equation}%
and the order of the Hankel functions $H_{\nu }^{(1,2)}(z)$ is related to
the mass of the field by
\begin{equation}
\nu =\left[ D^{2}/4-D(D+1)\xi -m^{2}\alpha ^{2}\right] ^{1/2}.  \label{nu}
\end{equation}%
This parameter is either real and nonnegative or purely imaginary.

An important step in formulating quantum field theory in a curved spacetime
is the choice of vacuum. Different choices of the coefficients in the linear
combination of the Hankel functions correspond to different choices of the
vacuum state. dS spacetime has maximal symmetry and it is natural to choose
a vacuum state having the same symmetry. It is well known that in dS
spacetime there exists a one-parameter family of maximally symmetric vacuum
states, which have been dubbed the $\alpha $-vacua (see Ref. \cite{Alle85}
and references therein). Among them the Bunch-Davies vacuum \cite{Bunc78} is
the only one with the Hadamard singularity structure.

Here we assume that the scalar field is prepared in the Bunch-Davies vacuum.
The mode functions realizing this state read:%
\begin{eqnarray}
\varphi _{\sigma }^{(+)}(x) &=&C_{\sigma }^{(+)}\eta ^{D/2}H_{\nu
}^{(1)}(k\eta )e^{i\mathbf{k}_{p}\cdot \mathbf{x}_{p}\mathbf{+ik}_{q}\cdot
\mathbf{x}_{q}},  \notag \\
\varphi _{\sigma }^{(-)}(x) &=&C_{\sigma }^{(-)}\eta ^{D/2}H_{\nu ^{\ast
}}^{(2)}(k\eta )e^{i\mathbf{k}_{p}\cdot \mathbf{x}_{p}\mathbf{+ik}_{q}\cdot
\mathbf{x}_{q}}.  \label{phipl}
\end{eqnarray}%
In these expressions, for the components of the momentum along
uncompactified dimensions one has $-\infty <k_{l}<+\infty $, $l=1,\ldots ,p$%
, and the eigenvalues of the components along the compact dimensions are
quantized by the periodicity conditions (\ref{BC1}):%
\begin{equation}
k_{l}=\left( 2\pi n_{l}+\tilde{\alpha}_{l}\right) /L_{l},\quad n_{l}=0,\pm
1,\pm 2,\ldots ,  \label{klComp}
\end{equation}%
for $l=p+1,...,D$. Hence, the mode functions are specified by the set of
quantum numbers $\sigma =(\mathbf{k}_{p},\mathbf{n}_{q})$, where $\mathbf{n}%
_{q}=(n_{p+1},\ldots ,n_{D})$.

The coefficients $C_{\sigma }^{(\pm )}$ in (\ref{phipl}) are determined from
the orthonormalization condition%
\begin{equation}
\int d^{D}x\sqrt{|g|}\left[ \varphi _{\sigma }^{(s)}(x)\partial _{t}\varphi
_{\sigma ^{\prime }}^{(s^{\prime })\ast }(x)-\varphi _{\sigma ^{\prime
}}^{(s^{\prime })\ast }(x)\partial _{t}\varphi _{\sigma }^{(s)}(x)\right]
=i\delta _{\sigma \sigma ^{\prime }}\delta _{ss^{\prime }},  \label{NormCond}
\end{equation}%
with $\delta _{\sigma \sigma ^{\prime }}=\delta (\mathbf{k}_{p}-\mathbf{k}%
_{p}^{\prime })\prod\nolimits_{l=p+1}^{D}\delta _{n_{l}n_{l}^{\prime }}$ and
$g$ being the determinant of the metric tensor corresponding to the
line-element (\ref{ds2}). By using the Wronskian relation for the Hankel
functions, one finds%
\begin{equation}
|C_{\sigma }^{(\pm )}|^{2}=\frac{\alpha ^{1-D}e^{i(\nu -\nu ^{\ast })\pi /2}%
}{2^{p+2}\pi ^{p-1}V_{q}},  \label{NormCoef}
\end{equation}%
where the star stands for the complex conjugate and $V_{q}=L_{p+1}...L_{D}$
is the volume of the compact subspace.

Substituting the functions (\ref{phipl}) into the mode-sum formula (\ref{Had}%
), for the Hadamard function we get the representation
\begin{eqnarray}
G^{(1)}(x,x^{\prime }) &=&\frac{\left( \eta \eta ^{\prime }\right)
^{D/2}e^{i(\nu -\nu ^{\ast })\pi /2}}{2^{p+2}\pi ^{p-1}V_{q}\alpha ^{D-1}}%
\int d\mathbf{k}_{p}\,e^{i\mathbf{k}_{p}\cdot \Delta \mathbf{x}_{p}}\sum_{%
\mathbf{n}_{q}}e^{i\mathbf{k}_{q}\cdot \Delta \mathbf{x}_{q}}  \notag \\
&&\times \lbrack H_{\nu }^{(1)}(k\eta )H_{\nu ^{\ast }}^{(2)}(k\eta ^{\prime
})+H_{\nu ^{\ast }}^{(2)}(k\eta )H_{\nu }^{(1)}(k\eta ^{\prime })],
\label{Had2}
\end{eqnarray}%
where $\Delta \mathbf{x}_{p}\mathbf{=x}_{p}-\mathbf{x}_{p}^{\prime }$, $%
\Delta \mathbf{x}_{q}\mathbf{=x}_{q}-\mathbf{x}_{q}^{\prime }$. In (\ref%
{Had2}) and in the formulas below we use the notation
\begin{equation}
\sum_{\mathbf{n}}=\sum_{n_{1}=-\infty }^{+\infty }\cdots \sum_{n_{l}=-\infty
}^{+\infty },  \label{Not}
\end{equation}%
for $\mathbf{n}=(n_{1},\ldots ,n_{l})$. Note that for a scalar field in dS
spacetime without compact dimensions ($p=D$), the corresponding two-point
functions contain infrared divergences for ${\mathrm{Re\,}}\nu \geq D/2$ and
in this region the Bunch-Davies vacuum is not a physically realizable state.
In particular, this is the case for a minimally coupled massless scalar
field. These divergences come from the singular behavior of the Hankel
functions at the origin. For the topology under consideration, assuming that
$\sum_{l=p+1}^{D}$ $\tilde{\alpha}_{l}^{2}\neq 0$ and $|\tilde{\alpha}%
_{l}|<\pi $, the momentum $k$ has a nonzero minimum value $k_{\mathrm{\min }%
}=\sqrt{\sum_{l=p+1}^{D}\tilde{\alpha}_{l}^{2}/L_{l}^{2}}$. In this case the
two-point function (\ref{Had2}) contains no infrared divergences and the
Bunch-Davies vacuum is a physically realizable state for all values of the
parameter $\nu $.

For the further evaluation of the Hadamard function, given by (\ref{Had2}),
we apply to the series over $n_{r}$, $p+1\leq r\leq D$, the Abel-Plana-type
summation formula \cite{Beze08,Bell09}%
\begin{eqnarray}
&&\frac{2\pi }{L_{r}}\sum_{n_{r}=-\infty }^{\infty }g(k_{r})f(\left\vert
k_{r}\right\vert )=\int_{0}^{\infty }dz[g(z)+g(-z)]f(z)  \notag \\
&&\qquad +i\int_{0}^{\infty }dz\,[f(iz)-f(-iz)]\sum_{\lambda =\pm 1}\frac{%
g(i\lambda z)}{e^{zL_{r}+i\lambda \tilde{\alpha}_{r}}-1},  \label{AbelPlan1}
\end{eqnarray}%
with $k_{r}$ defined by (\ref{klComp}). In the special case $g(z)=1$ and $%
\tilde{\alpha}_{r}=0$ this formula is reduced to the standard Abel-Plana
formula (for generalizations of the Abel-Plana formula see \cite{SahaBook}).
For the series in (\ref{Had2}) one has $g(z)=e^{iz\Delta x^{r}}$ and the
function $f(z)$ is given by the expression in the square brackets. It can be
seen that the contribution of the first integral in the right-hand side of (%
\ref{AbelPlan1}) to the Hadamard function coincides with the corresponding
Hadamard function for the topology $R^{p+1}\times (S^{1})^{q-1}$ with the
lengths of the compact dimensions $(L_{p+1},\ldots ,L_{r-1},L_{r+1},\ldots
,L_{D})$.

In the part corresponding to the contribution of the second integral in (\ref%
{AbelPlan1}) we use the expansion $e^{y}-1=\sum_{n=1}^{\infty }e^{-ny}$.
After some intermediate calculations, the Hadamard function is presented in
the form%
\begin{eqnarray}
G^{(1)}(x,x^{\prime }) &=&\frac{4\left( \eta \eta ^{\prime }\right)
^{D/2}L_{r}}{\left( 2\pi \right) ^{(p+3)/2}V_{q}\alpha ^{D-1}}\sum_{\mathbf{n%
}_{q-1}^{(r)}}e^{i\mathbf{k}_{q-1}^{(r)}\cdot \Delta \mathbf{x}%
_{q-1}}\int_{0}^{\infty }dz\,z  \notag \\
&&\times \left[ I_{-\nu }(\eta z)K_{\nu }(\eta ^{\prime }z)+K_{\nu }(\eta
z)I_{\nu }(\eta ^{\prime }z)\right] \sum_{n=-\infty }^{\infty }e^{-in\alpha
_{r}}  \notag \\
&&\times \frac{f_{(p-1)/2}(\sqrt{z^{2}+u_{r}^{2}}\sqrt{\left( \Delta
x^{r}+nL_{r}\right) ^{2}+|\Delta \mathbf{x}_{p}|^{2}})}{[\left( \Delta
x^{r}+nL_{r}\right) ^{2}+|\Delta \mathbf{x}_{p}|^{2}]^{(p-1)/2}},
\label{Had3}
\end{eqnarray}%
where $I_{\nu }(y)$ and $K_{\nu }(y)$ are the modified Bessel functions, $%
\mathbf{n}_{q-1}^{(r)}=(n_{p+1},\ldots ,n_{r-1},n_{r+1},\ldots ,n_{D})$, $%
\mathbf{k}_{q-1}^{(r)}=(k_{p+1},\ldots ,k_{r-1},k_{r+1},\ldots ,k_{D})$ and%
\begin{equation}
u_{r}^{2}=|\mathbf{k}_{q-1}^{(r)}|^{2}=\sum_{l=p+1,\neq r}^{D}\left( 2\pi
n_{l}+\tilde{\alpha}_{l}\right) ^{2}/L_{l}^{2}.  \label{ur}
\end{equation}%
Here and in what follows we use the notation%
\begin{equation}
f_{\mu }(z)=z^{\mu }K_{\mu }(z).  \label{fmu}
\end{equation}%
The representation (\ref{Had3}) of the Hadamard function is valid for ${%
\mathrm{Re}\,\nu <1}$. The $n=0$ term in this formula is the Hadamard
function for the topology $R^{p+1}\times (S^{1})^{q-1}$ with the lengths of
the compact dimensions $(L_{p+1},\ldots ,L_{r-1},L_{r+1},\ldots ,L_{D})$.
The remaining part vanishes in the limit $L_{r}\rightarrow \infty $ and it
is induced by the compactification of the $r$th dimension. The expression (%
\ref{Had3}) can be used for the investigation of the VEVs for various
physical observables. Here we consider the VEV of the current density.

\section{Current density for a scalar field}

\label{Sec:Sc}

Having the Hadamard function, we can evaluate the expectation value for the
current density by using the formula (\ref{jscH}). It can be seen that the
VEVs of the charge density ($\mu =0$) and of the components of the current
density along uncompactified dimensions vanish:
\begin{equation}
\left\langle j_{\mu }\right\rangle =0,\;\mu =0,1,\ldots ,p.  \label{Charge}
\end{equation}%
The latter is a direct consequence of the problem symmetry. For a noncompact
dimension $x^{l}$ the problem is symmetric under the reflection $%
x^{l}\rightarrow -x^{l}$. The presence of the nonzero current density along $%
x^{l}$ would break this symmetry.

For the VEV of the current density along the $r$th compact dimension, from (%
\ref{jscH}) and (\ref{Had3}) we find%
\begin{eqnarray}
\left\langle j^{r}\right\rangle &=&\frac{8e\alpha (\eta /\alpha )^{D+2}}{%
\left( 2\pi \right) ^{(p+3)/2}V_{q}L_{r}^{p-1}}\int_{0}^{\infty }dz\,z\left[
I_{-\nu }(\eta z)+I_{\nu }(\eta z)\right] K_{\nu }(\eta z)  \notag \\
&&\times \sum_{n=1}^{\infty }\frac{\sin (n\tilde{\alpha}_{r})}{n^{p}}\sum_{%
\mathbf{n}_{q-1}^{(r)}}f_{(p+1)/2}(nL_{r}\sqrt{z^{2}+u_{r}^{2}}).
\label{jscr}
\end{eqnarray}%
This VEV is an even periodic function of the phases $\tilde{\alpha}_{l}$, $%
l\neq r$, with the period $2\pi $, and it is an odd periodic function of the
phase $\tilde{\alpha}_{r}$ with the same period. In particular, the current
density is a periodic function of the magnetic flux with the period of the
flux quantum. In the absence of the gauge field the VEV of the current
density vanishes for special cases of twisted ($\tilde{\alpha}_{r}=\pi $)
and untwisted ($\tilde{\alpha}_{r}=0$) fields. In the special case of a
single compact dimension one has $p=D-1$, $q=1$, and the general formula
takes the form
\begin{equation}
\left\langle j^{r}\right\rangle =\frac{8e\alpha (\eta /\alpha )^{D+2}}{%
\left( 2\pi \right) ^{D/2+1}L_{r}^{D-1}}\int_{0}^{\infty }dz\,\,z\left[
I_{-\nu }(\eta z)+I_{\nu }(\eta z)\right] K_{\nu }(\eta z)\sum_{n=1}^{\infty
}\frac{\sin (n\tilde{\alpha}_{r})}{n^{D-1}}f_{D/2}(nL_{r}z).
\label{jscrSing}
\end{equation}

From the covariant conservation equation $\nabla _{\mu }\left\langle j^{\mu
}\right\rangle =0$ it follows that the charge flux through the $(D-1)$%
-dimensional spatial hypersurface $x^{r}=\mathrm{const}$ is determined by
the quantity $n_{r}\left\langle j^{r}\right\rangle $, where $n_{r}=\sqrt{%
|g_{rr}|}=\alpha /\eta $ is the normal to the hypersurface. Now, from (\ref%
{jscr}) we see that this quantity depends on the time coordinate and on the
lengths of the compact dimensions in the form of the ratio $L_{l}/\eta $:%
\begin{equation}
n_{r}\left\langle j^{r}\right\rangle =\alpha ^{-D}f(L_{p+1}/\eta ,\ldots
,L_{D}/\eta ).  \label{nrjr}
\end{equation}%
By taking into account that $L_{l}^{\mathrm{(p)}}=\alpha L_{l}/\eta $ is the
proper length of the compact dimension, we see that the ratio $L_{l}/\eta $
is the proper length of the $l$th compact dimension measured in the units of
the dS curvature scale $\alpha $.

For a conformally coupled massless scalar field one has $m=0$, $\xi
=(D-1)/(4D)$ and, hence, $\nu =1/2$. By taking into account the
corresponding expressions for the modified Bessel function, we see that $%
\left[ I_{-\nu }(x)+I_{\nu }(x)\right] K_{\nu }(x)=1/x$. The integral in (%
\ref{jscr}) is evaluated by using the formula
\begin{equation}
\int_{0}^{\infty }du\,f_{\mu }(a\sqrt{u^{2}+b^{2}})=\frac{1}{a}\sqrt{\frac{%
\pi }{2}}f_{\mu +1/2}(ab),  \label{IntForm1}
\end{equation}%
and for the VEV of the current density we get%
\begin{equation}
\left\langle j^{r}\right\rangle =\frac{4e(\eta /\alpha )^{D+1}}{\left( 2\pi
\right) ^{p/2+1}V_{q}L_{r}^{p}}\sum_{n=1}^{\infty }\frac{\sin (n\tilde{\alpha%
}_{r})}{n^{p+1}}\sum_{\mathbf{n}_{q-1}^{(r)}}f_{p/2+1}(nL_{r}u_{r}).
\label{jscConf}
\end{equation}%
In this case, the current density is related to the corresponding result in
the Minkowski spacetime (see Ref. \cite{Beze12}) with compact dimensions of
the lengths $(L_{p+1},\ldots ,L_{D})$ by the equation $\left\langle
j^{r}\right\rangle =(\eta /\alpha )^{D+1}\left\langle j^{r}\right\rangle ^{%
\mathrm{(M)}}$.

Let us consider the Minkowskian limit for the general expression of the
current density given by (\ref{jscr}). This corresponds to the limit $\alpha
\rightarrow \infty $ for a fixed value of $t$. In this limit one has $\nu
\approx i\beta $, with $\beta =m\alpha \gg 1$ and $\eta \approx \alpha -t$.
By using the uniform asymptotic expansions for the modified Bessel functions
for imaginary values of the order with large modulus (see, for example, Ref.~%
\cite{Duns90}), we can see that for $x<1$ one has%
\begin{equation}
\left[ I_{i\beta }(\beta x)+I_{-i\beta }(\beta x)\right] K_{i\beta }(\beta
x)\sim \frac{1}{\beta }\cos [2\beta f(x)],  \label{BesComb1}
\end{equation}
where
\begin{equation}
f(x)=\ln \left( \frac{1+\sqrt{1-x^{2}}}{x}\right) -\sqrt{1-x^{2}}.
\label{fz}
\end{equation}%
In the case $x>1$ the leading term is given by%
\begin{equation}
\left[ I_{i\beta }(\beta x)+I_{-i\beta }(\beta x)\right] K_{i\beta }(\beta
x)\sim \frac{1}{\beta \sqrt{x^{2}-1}}.  \label{BesComb2}
\end{equation}%
From these expressions it follows that the dominant contribution to the
current density in (\ref{jscr}) comes from the integration range $z>m$. In
this range, by using (\ref{BesComb2}) and the integration formula (\ref%
{IntForm1}), to the leading order we find%
\begin{equation}
\left\langle j^{r}\right\rangle \approx \frac{4eL_{r}^{-p}}{\left( 2\pi
\right) ^{p/2+1}V_{q}}\sum_{n=1}^{\infty }\frac{\sin (n\tilde{\alpha}_{r})}{%
n^{p+1}}\sum_{\mathbf{n}_{q-1}^{(r)}}f_{p/2+1}(nL_{r}\sqrt{u_{r}^{2}+m^{2}}).
\label{jscMink}
\end{equation}%
The expression in the right-hand side coincides with the VEV of the current
density in the Minkowski bulk with toroidally compactified dimensions \cite%
{Beze12}. For this background geometry, the finite temperature corrections
to the expectation value of the current density have been investigated in
Ref. \cite{Beze12}. In this reference, we have also derived an alternative
expression for the current density in the Minowskian bulk by using the zeta
function approach. In the case of a massless field, the expression in the
right-hand side of (\ref{jscMink}) reduces to the result which follows from (%
\ref{jscConf}).

Now we turn to the investigation of the current density in the asymptotic
regions of the ratio $L_{r}/\eta $. As we have mentioned before, this
quantity is the ratio of the comoving length of the compact dimension to the
curvature radius of dS spacetime. For small values of this ratio, $%
L_{r}/\eta \ll 1$, we introduce in (\ref{jscr}) a new integration variable $%
y=L_{r}z$. By taking into account that for large values of $x$ one has $%
\left[ I_{-\nu }(x)+I_{\nu }(x)\right] K_{\nu }(x)\approx 1/x$, we find that
to the leading order $\left\langle j^{r}\right\rangle $ coincides with the
corresponding result for a conformally coupled massless field (see (\ref%
{jscConf})):%
\begin{equation}
\left\langle j^{r}\right\rangle \approx (\eta /\alpha )^{D+1}\left\langle
j^{r}\right\rangle ^{\mathrm{(M)}},\;L_{r}/\eta \ll 1.  \label{jscearly}
\end{equation}%
For a fixed value of the ratio $L_{r}/\alpha $, the limit under
consideration corresponds to the early stages of the cosmological expansion,
$t\rightarrow -\infty $, and the current density behaves like $\exp
[-(D+1)t/\alpha ]$. Note that in the limit $L_{r}/\eta \ll 1$, the leading
term in the VEV of the quantity $n_{r}\left\langle j^{r}\right\rangle $
coincides with the corresponding quantity in the Minkowskian bulk, if we
replace the lengths of the compact dimensions $L_{l}$ by the proper lengths $%
\alpha L_{l}/\eta $. In this limit, the effects induced by the curvature of
the background spacetime are small.

In the limit of large values of the proper length of the $r$th compact
dimension, compared with the dS curvature scale, one has $\eta /L_{r}\ll 1$.
In this case, we introduce in (\ref{jscr}) a new integration variable $%
y=L_{r}z$ and then expand the integrand by using the formulas for the
modified Bessel functions for small values of the argument. Two cases should
be considered separately. For positive values of the parameter $\nu $, after
the integration over $y$ by using the formula from \cite{Prud86}, to the
leading order we find%
\begin{equation}
\left\langle j^{r}\right\rangle \approx e\frac{2^{\nu -(p-1)/2}(\eta
/L_{r})^{D-2\nu +2}\Gamma (\nu )}{\pi ^{(p+3)/2}\alpha
^{D+1}V_{q-1}L_{r}^{p-D}}\sum_{n=1}^{\infty }\frac{\sin (n\tilde{\alpha}_{r})%
}{n^{p-2\nu +2}}\sum_{\mathbf{n}_{q-1}^{(r)}}f_{(p+3)/2-\nu }(nL_{r}u_{r}),
\label{jsclate}
\end{equation}%
for $\eta /L_{r}\ll 1$. In the case of a conformally coupled massless scalar
field $\nu =1/2$ and (\ref{jsclate}) reduces to the exact result given by (%
\ref{jscConf}). For a fixed value of $L_{r}/\alpha $, the limit under
consideration corresponds to late stages of the cosmological evolution, $%
t\rightarrow +\infty $, and the current density is suppressed by the factor $%
\exp [-(D-2\nu +2)t/\alpha ]$. Note that formula (\ref{jsclate}) also
describes the asymptotic behavior for the current density in the strong
curvature regime corresponding to small values of the parameter $\alpha $.
In the model with a single compact dimension ($p=D-1$) the asymptotic
formula (\ref{jsclate}) takes the form%
\begin{equation}
\left\langle j^{r}\right\rangle \approx \frac{2e\Gamma (\nu )\Gamma
(D/2+1-\nu )\eta }{\pi ^{D/2+1}\alpha ^{D+1}(L_{r}/\eta )^{D-2\nu +1}}%
\sum_{n=1}^{\infty }\frac{\sin (n\tilde{\alpha}_{r})}{n^{D-2\nu +1}}.
\label{jsclateSing}
\end{equation}%
The sum in the latter expression is expressed in terms of the Hurwitz zeta
function $\zeta (s,x)$.

In the same limit, $\eta /L_{r}\ll 1$, and for pure imaginary values of the
parameter $\nu $, by taking into account that for small values of $x$ one has%
\begin{equation}
\left[ I_{-\nu }(x)+I_{\nu }(x)\right] K_{\nu }(x)\approx {\mathrm{Re}}\left[
\frac{(2/x)^{2i|\nu |}\Gamma (i|\nu |)}{\Gamma (1-i|\nu |)}\right] ,
\label{BessSmall}
\end{equation}%
for the current density we find the following asymptotic behavior%
\begin{equation}
\left\langle j^{r}\right\rangle \approx \frac{8e\alpha
B_{s}e^{-(D+2)t/\alpha }}{\left( 2\pi \right) ^{(p+3)/2}V_{q}L_{r}^{p+1}}%
\cos [2|\nu |t/\alpha +2|\nu |\ln (L_{r}/\alpha )+\phi _{s}].
\label{jsclate2}
\end{equation}%
In this formula, $B_{s}$ and $\phi _{s}$ are defined by the relation
\begin{equation}
B_{s}e^{i\phi _{s}}=2^{i|\nu |}\Gamma (i|\nu |)\sum_{n=1}^{\infty }\frac{%
\sin (n\tilde{\alpha}_{r})}{n^{p+2-2i|\nu |}}\sum_{\mathbf{n}%
_{q-1}^{(r)}}f_{(p+3)/2-i|\nu |}(nL_{r}u_{r}).  \label{Bphi0}
\end{equation}%
Hence, in the case under consideration, at late stages of the cosmological
evolution the current density is suppressed by the factor $e^{-(D+2)t/\alpha
}$ and the damping of the corresponding VEV has an oscillatory nature.

For a fixed value of the time coordinate $t$, the condition $\eta /L_{r}\ll
1 $ corresponds to the length of the $r$th compact dimension much larger
than the dS curvature scale, $L_{r}\gg \alpha $. Now for the values of the
curvature scale $\alpha \gtrsim m^{-1}$ one also has $L_{r}\gg m^{-1}$ and
the limit we have discussed corresponds to the length of the compact
dimension much larger than the Compton wavelength of the field quanta. As it
is seen from the formula (\ref{jsclateSing}), in this range the decay of the
current density is a power-law as a function of $mL_{r}$. This is in
contrast to the case of the Minkowski bulk, where, as it can be seen from (%
\ref{jscMink}), the VEV of the current density is suppressed exponentially.
For example, in the case of a single compact dimension ($q=1$) one has%
\begin{equation*}
\left\langle j^{r}\right\rangle ^{\mathrm{(M)}}\approx \frac{2em^{D}\sin (%
\tilde{\alpha}_{r})}{\left( 2\pi \right) ^{D/2}(mL_{r})^{D/2}}e^{-mL_{r}},
\end{equation*}%
for $mL_{r}\gg 1$.

\begin{figure}[tbph]
\begin{center}
\epsfig{figure=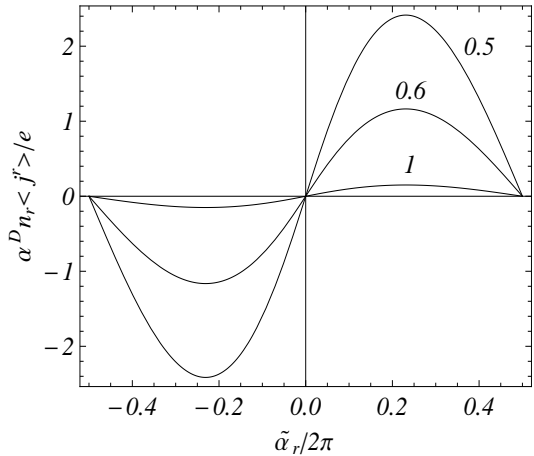,width=7.cm,height=6.cm}
\end{center}
\caption{The quantity $\protect\alpha ^{D}n_{r}\left\langle
j^{r}\right\rangle /e$ as a function of the phase in the quasiperiodicity
condition for a conformally coupled scalar field in the $D=4$ model with a
single compact dimension. The graphs are plotted for $\protect\alpha m=0.25$
and the numbers near the curves are the corresponding values for the ratio $%
L_{r}/\protect\eta $. }
\label{fig1}
\end{figure}

In figure \ref{fig1} we display the quantity $\alpha ^{D}n_{r}\left\langle
j^{r}\right\rangle /e$ as a function of the phase in the quasiperiodicity
condition for a conformally coupled scalar field in the $D=4$
Kaluza-Klein-type model with a single compact dimension $x^{r}=x^{4}$ ($p=3$%
). The graphs are plotted for $\alpha m=0.25$ and the numbers near the
curves are the corresponding values for the ratio $L_{r}/\eta $. Let us
recall that the current density is a periodic function of $\tilde{\alpha}%
_{r} $ with the period equal $2\pi $. For a field with the periodic boundary
condition ($\alpha _{r}=0$) one has $\tilde{\alpha}_{r}/2\pi =-\Phi
_{r}/\Phi _{0}$.

\begin{figure}[tbph]
\begin{center}
\begin{tabular}{cc}
\epsfig{figure=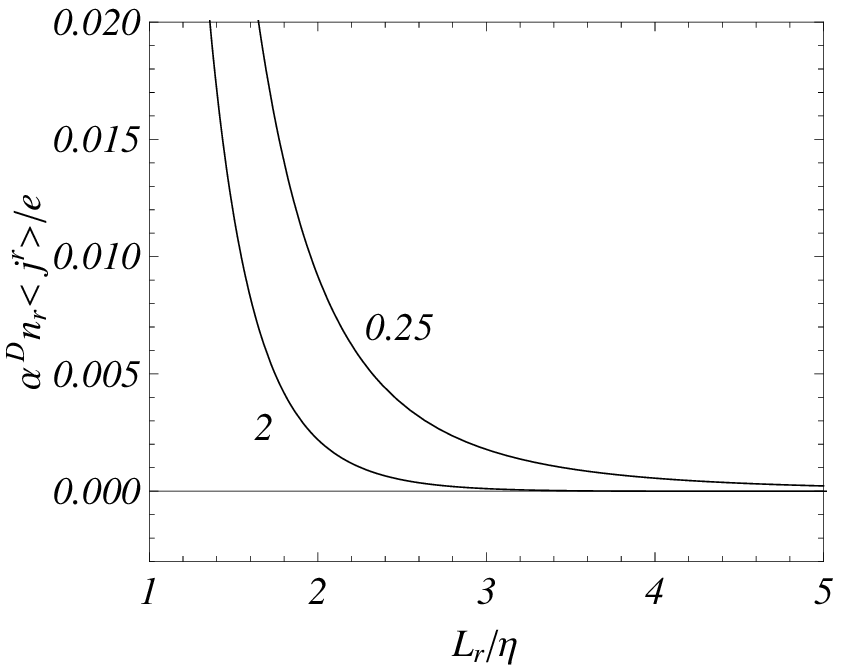,width=7.cm,height=6.cm} & \quad %
\epsfig{figure=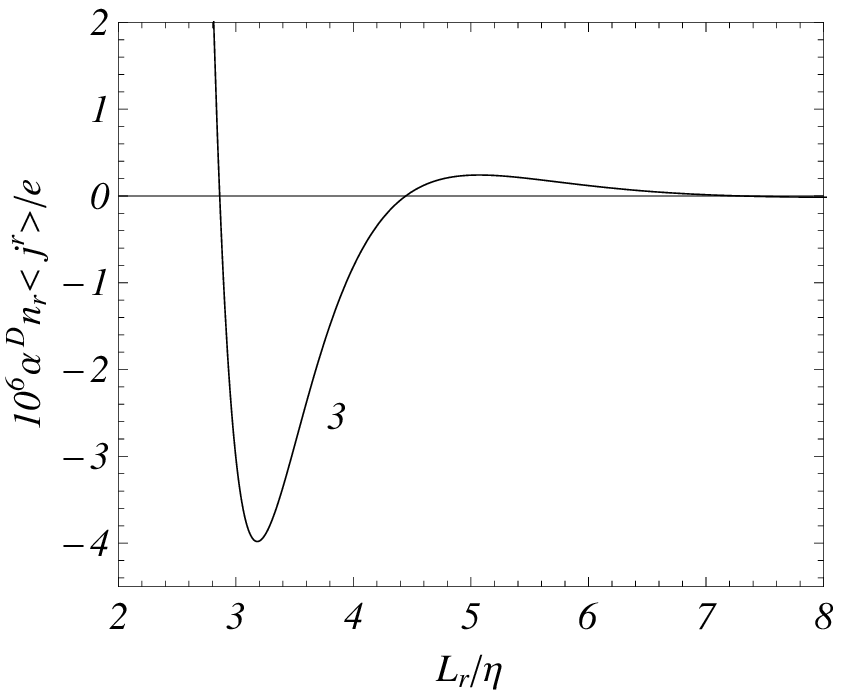,width=7.cm,height=6cm}%
\end{tabular}%
\end{center}
\caption{The current density for a conformally coupled scalar field in the
model $D=4$ and $p=3$ as a function of the proper length of the compact
dimension measured in units of the curvature scale $\protect\alpha $. The
numbers near the curves are the values of the parameter $m\protect\alpha $,
and for the phase in the periodicity condition we have taken $\tilde{\protect%
\alpha}_{r}=\protect\pi /2$. }
\label{fig2}
\end{figure}

For the same model $D=4$, $p=3$, in figure \ref{fig2}, we have plotted the
quantity $\alpha ^{D}n_{r}\left\langle j^{r}\right\rangle $ for a
conformally coupled scalar field versus the ratio $L_{r}/\eta $. The latter
is the proper length of the compact dimension measured in units of the dS
curvature scale $\alpha $. In numerical evaluations we have taken $\tilde{%
\alpha}_{r}=\pi /2$ and the numbers near the curves correspond to the values
of $m\alpha $. For small values of $L_{r}/\eta $, from the asymptotic
analysis given above it follows that $n_{r}\left\langle j^{r}\right\rangle
\propto (L_{r}/\eta )^{-D}$. As it has been explained before, depending on
the parameter $\nu $, for large values of $L_{r}/\eta $ two different
regimes arise with monotonic (for positive values of $\nu $) and oscillatory
(for imaginary values of $\nu $) damping of the current density. Note that,
for a conformally coupled field, $\nu $ is real for $m\alpha =0.25$ and
imaginary for $m\alpha =2,3$. In order to display the oscillatory behavior
of the damping, on the right panel of figure \ref{fig2} we have plotted the
graph for $m\alpha =3$. The value of the ratio $L_{r}/\eta $ corresponding
to the first zero of the current density decreases with increasing the value
of $m\alpha $. For the first two zeros of the current density one has $%
L_{r}/\eta =3.88,8.29$ and $L_{r}/\eta =2.87,4.45$ in the cases $m\alpha =2$
and $m\alpha =3$, respectively.

\section{Fermionic current}

\label{Sec:Ferm}

In this section we consider the VEV\ of the current density for a fermionic
field $\psi (x)$ on the background of $(D+1)$-dimensional dS spacetime with
spatial topology $R^{p}\times (S^{1})^{q}$. As before, the line-element is
taken in the form (\ref{ds2}). Assuming the presence of a classical gauge
field $A_{\mu }$, the dynamics of the Dirac spinor field is governed by the
equation
\begin{equation}
i\gamma ^{\mu }D_{\mu }\psi -m\psi =0\ ,\;D_{\mu }=\partial _{\mu }+\Gamma
_{\mu }+ieA_{\mu }.  \label{Direq}
\end{equation}%
The Dirac matrices and the spin connection $\Gamma _{\mu }$ are expressed in
terms of the flat-space Dirac matrices $\gamma ^{(a)}$ by the relations
\begin{equation}
\gamma ^{\mu }=e_{(a)}^{\mu }\gamma ^{(a)},\;\Gamma _{\mu }=\frac{1}{4}%
\gamma ^{(a)}\gamma ^{(b)}e_{(a)}^{\nu }e_{(b)\nu ;\mu }\ ,  \label{Gammamu}
\end{equation}%
where the semicolon means the covariant derivative of vector fields and $%
e_{(a)}^{\mu }$ are the tetrads defined by $e_{(a)}^{\mu }e_{(b)}^{\nu }\eta
^{ab}=g^{\mu \nu }$. For the background under consideration the tetrads can
be taken in the form%
\begin{equation}
e_{\mu }^{(0)}=\delta _{\mu }^{0},\;e_{\mu }^{(a)}=e^{t/\alpha }\delta _{\mu
}^{a},\;a=1,2,\ldots ,D.  \label{Tetrad}
\end{equation}%
In a $(D+1)$-dimensional spacetime the Dirac matrices are $N\times N$
matrices with $N=2^{[(D+1)/2]}$, where the square brackets denote the
integer part of the enclosed expression. Here we take the flat spacetime
matrices in the Dirac representation
\begin{equation}
\gamma ^{(0)}=\left(
\begin{array}{cc}
1 & 0 \\
0 & -1%
\end{array}%
\right) ,\;\gamma ^{(a)}=\left(
\begin{array}{cc}
0 & \sigma _{a} \\
-\sigma _{a}^{+} & 0%
\end{array}%
\right) .  \label{gam0l}
\end{equation}%
with $a=1,2,\ldots ,D$.

We need also to specify the periodicity conditions for the fermionic field
along compact dimensions. Here these conditions will be taken in the form%
\begin{equation}
\psi (t,\mathbf{x}_{p},\mathbf{x}_{q}+L_{l}\mathbf{e}_{l})=e^{i\alpha
_{l}}\psi (t,\mathbf{x}_{p},\mathbf{x}_{q}),  \label{bc}
\end{equation}%
with constant phases $\alpha _{l}$. Although we have used the same notation
for the phases in the boundary conditions for scalar and fermionic fields,
of course, in general they can be different. The fermionic condensate and
the VEV of the energy-momentum tensor in the special cases of untwisted ($%
\alpha _{l}=0$) and twisted ($\alpha _{l}=\pi $) fermionic fields and in the
absence of a gauge field have been investigated in \cite{Saha08,Beze09}.

Assuming a constant gauge field $A_{\mu }$, we can exclude it from the field
equation by the gauge transformation $A_{\mu }^{\prime }=A_{\mu }+\partial
_{\mu }\omega $, $\psi ^{\prime }(x)=e^{-ie\omega }\psi (x)$ with the
function $\omega $ defined in (\ref{Gauge}). For the new field we have the
quasiperiodicity condition%
\begin{equation}
\psi ^{\prime }(t,\mathbf{x}_{p},\mathbf{x}_{q}+L_{l}\mathbf{e}_{l})=e^{i%
\tilde{\alpha}_{l}}\psi ^{\prime }(t,\mathbf{x}_{p},\mathbf{x}_{q}),
\label{bcf2}
\end{equation}
with the new phase defined by (\ref{alf1}). Similarly to the case for a
scalar field, we will work in terms of the gauge transformed field omitting
the prime.

The VEV of the current density for the fermionic field can be expressed in
terms of the two-point function $S_{rs}^{(1)}(x,x^{\prime })=\langle 0|[\psi
_{r}(x),\bar{\psi}_{s}(x^{\prime })]|0\rangle $, where $r$ and $s$ are
spinor indices. The expression for the VEV reads%
\begin{equation}
\langle j^{\mu }(x)\rangle \equiv \langle 0|j^{\mu }(x)|0\rangle =-\frac{e}{2%
}\mathrm{Tr}(\gamma ^{\mu }S^{(1)}(x,x)).  \label{VEVj}
\end{equation}%
Expanding the field operator in terms of a complete set of solutions to the
Dirac equation, $\{\psi _{\sigma }^{(\pm )}(x)\}$, the following mode-sum
formula can be obtained:
\begin{equation}
\langle j^{\mu }\rangle =-\frac{e}{2}\sum_{\sigma }\sum_{s=\pm }s\bar{\psi}%
_{\sigma }^{(s)}(x)\gamma ^{\mu }\psi _{\sigma }^{(s)}(x).  \label{jfmode}
\end{equation}%
In order to regularize the divergent expression in the right-hand side we
will assume the presence of a cutoff function without writing it explicitly.
The special form of this function will not be important for the further
discussion.

For the problem under consideration, the complete set of fermionic mode
functions can be found in a way similar to that used in \cite{Beze09} for
special cases of twisted and untwisted fields. These mode function are
presented in the form
\begin{eqnarray}
\psi _{\sigma }^{(+)} &=&A_{\sigma }^{(+)}\eta ^{(D+1)/2}e^{i\mathbf{k}\cdot
\mathbf{x}}\left(
\begin{array}{c}
H_{1/2-i\alpha m}^{(1)}(k\eta )w_{\chi }^{(+)} \\
-i(\mathbf{n}\cdot \boldsymbol{\sigma }^{+})H_{-1/2-i\alpha m}^{(1)}(k\eta
)w_{\chi }^{(+)}%
\end{array}%
\right) ,  \label{psisig+} \\
\psi _{\sigma }^{(-)} &=&A_{\sigma }^{(-)}\eta ^{(D+1)/2}e^{i\mathbf{k}\cdot
\mathbf{x}}\left(
\begin{array}{c}
-i(\mathbf{n}\cdot \boldsymbol{\sigma })H_{-1/2+i\alpha m}^{(2)}(k\eta
)w_{\chi }^{(-)} \\
H_{1/2+i\alpha m}^{(2)}(k\eta )w_{\chi }^{(-)}%
\end{array}%
\right) ,  \label{psisig-}
\end{eqnarray}%
where $\mathbf{n}=\mathbf{k}/k$, $\boldsymbol{\sigma }=(\sigma _{1},\ldots
,\sigma _{D})$, and $\sigma =(\mathbf{k},\chi )$. In (\ref{psisig+}) and (%
\ref{psisig-}), $w_{\chi }^{(\pm )}$, $\chi =1,\ldots ,N/2$, are one-column
matrices having $N/2$ rows with the elements $w_{\chi l}^{(+)}=\delta
_{l\chi }$ and $w_{\chi }^{(-)}=iw_{\chi }^{(+)}$. For the normalization
coefficients one has%
\begin{equation}
|A_{\sigma }^{(\pm )}|^{2}=\frac{ke^{\pi \alpha m}}{2^{p+2}\pi
^{p-1}V_{q}\alpha ^{D}}.  \label{Abet}
\end{equation}%
Similarly to the case of a scalar field, the eigenvalues for the components
of the momentum along compact dimensions are given by the expressions (\ref%
{klComp}). Note that for a massless field we have the standard conformal
relation $\psi _{\beta }^{(\pm )}=(\eta /\alpha )^{(D+1)/2}\psi _{\beta }^{%
\mathrm{(M)}(\pm )}$ between the mode functions defining the Bunch-Davies
vacuum in dS spacetime and the corresponding mode functions $\psi _{\beta }^{%
\mathrm{(M)}(\pm )}$ in the Minkowski spacetime with spatial topology $%
R^{p}\times (S^{1})^{q}$.

Substituting the mode-functions into the mode-sum formula (\ref{jfmode}), it
can be seen that the VEVs of the charge density and of the components along
uncompactified dimensions vanish: $\langle j^{\mu }\rangle =0$ for $\mu
=0,1,\ldots ,p$. For the component of the current density along the $r$th
compact dimension we get the expression%
\begin{eqnarray}
\langle j^{r}\rangle &=&\frac{ie\pi ^{1-p/2}\eta ^{D+2}e^{\pi \alpha m}N}{%
2^{p+2}V_{q}\alpha ^{D+1}\Gamma (p/2)}\int_{0}^{\infty
}dk_{p}\,k_{p}^{p-1}\sum_{\mathbf{n}_{q}}k_{r}  \notag \\
&&\times \lbrack H_{-1/2-i\alpha m}^{(1)}(k\eta )H_{1/2+i\alpha
m}^{(2)}(k\eta )-H_{1/2-i\alpha m}^{(1)}(k\eta )H_{-1/2+i\alpha
m}^{(2)}(k\eta )],  \label{jrf}
\end{eqnarray}%
where%
\begin{equation}
k^{2}=k_{p}^{2}+\sum_{l=p+1}^{D}(2\pi n_{l}+\tilde{\alpha}%
_{l})^{2}/L_{l}^{2}.  \label{k2f}
\end{equation}

Further evaluation of the current density is similar to that for the case of
the scalar field. Applying the summation formula (\ref{AbelPlan1}) to the
series over $n_{r}$ in (\ref{jrf}), one finds the expression

\begin{eqnarray}
\langle j^{r}\rangle &=&-\frac{4e\eta ^{D+2}L_{r}^{1-p}N}{(2\pi
)^{(p+3)/2}\alpha ^{D+1}V_{q}}\int_{0}^{\infty }dz\,z\,{\mathrm{Re}}\left\{ %
\left[ I_{-1/2-i\alpha m}(z\eta )+I_{1/2+i\alpha m}(z\eta )\right]
K_{1/2+i\alpha m}(z\eta )\right\}  \notag \\
&&\times \sum_{n=1}^{\infty }\frac{\sin (n\tilde{\alpha}_{r})}{n^{p}}\sum_{%
\mathbf{n}_{q-1}^{(r)}}f_{(p+1)/2}(nL_{r}\sqrt{z^{2}+u_{r}^{2}}),
\label{jrf2}
\end{eqnarray}%
where $u_{r}$ is defined by (\ref{ur}). Similarly to the case of the scalar
field, the fermionic current density $\langle j^{r}\rangle $ is an odd
periodic function of the phase $\tilde{\alpha}_{r}$ with the period $2\pi $
and it is an even periodic function of the phases $\tilde{\alpha}_{l}$, $%
l\neq r$, with the same period. In the absence of the gauge field the
current density vanishes for periodic and antiperiodic boundary conditions.
Similarly to the scalar case, the quantity $n_{r}\langle j^{r}\rangle $,
which describes the charge flux through the hypersurface $x^{r}=\mathrm{const%
}$, depends on $\eta $ and on the lengths of compact dimensions in the form
of the combination $L_{l}/\eta $. The latter is the proper length of the
compact dimension measured in units of $\alpha $.

For a massless fermionic field the modified Bessel functions are expressed
in terms of the exponential function and, after the integration by using the
formula (\ref{IntForm1}), from (\ref{jrf2}) we obtain the expression%
\begin{equation}
\langle j^{r}\rangle =-\frac{2e(\eta /\alpha )^{D+1}N}{(2\pi
)^{p/2+1}V_{q}L_{r}^{p}}\sum_{n=1}^{\infty }\frac{\sin (n\tilde{\alpha}_{r})%
}{n^{p+1}}\sum_{\mathbf{n}_{q-1}^{(r)}}f_{p/2+1}(nL_{r}u_{r}).  \label{jrfm0}
\end{equation}%
In this case one has the following relation between the fermionic and scalar
current densities:%
\begin{equation}
\langle j^{r}\rangle _{\mathrm{ferm}}=-N\langle j^{r}\rangle _{\mathrm{sc}%
}/2.  \label{jfjsc}
\end{equation}%
Here we assumed that the phases $\alpha _{l}$ are the same for both fields.
Note that in (\ref{jfjsc}), $\langle j^{r}\rangle _{\mathrm{sc}}$ is the
current density for a complex scalar field which is equivalent to two real
scalar fields.

The transition to the Minkowskian limit is similar to that for the case of a
scalar field. In this limit $\alpha \rightarrow \infty $ for a fixed $t$ and
$\eta \approx \alpha -t$. The dominant contribution to the integral in (\ref%
{jrf2}) comes from the integration region $z>m$. In this region we have
\begin{equation}
\left[ I_{-1/2-i\alpha m}(z\eta )+I_{1/2+i\alpha m}(z\eta )\right]
K_{1/2+i\alpha m}(z\eta )\approx \frac{1}{\alpha \sqrt{z^{2}-m^{2}}}.
\label{BessComb3}
\end{equation}%
After the integration with the help of (\ref{IntForm1}), to the leading
order we get the current density in the Minkowski spacetime $\langle
j^{r}\rangle \approx \langle j^{r}\rangle ^{\mathrm{(M)}}$ with
\begin{equation}
\langle j^{r}\rangle ^{\mathrm{(M)}}=-\frac{2eNL_{r}^{-p}}{\left( 2\pi
\right) ^{p/2+1}V_{q}}\sum_{n=1}^{\infty }\frac{\sin (n\tilde{\alpha}_{r})}{%
n^{p+1}}\sum_{\mathbf{n}_{q-1}^{(r)}}f_{p/2+1}(nL_{r}\sqrt{u_{r}^{2}+m^{2}}).
\label{jfM}
\end{equation}%
This result coincides with the formula derived in \cite{Bell10} (in the
notations of this reference $\tilde{\alpha}_{l}\rightarrow -2\pi \tilde{%
\alpha}_{l}$ and in \cite{Bell10} $A_{l}=\mathbf{A}_{l}$ in the definition
for $\tilde{\alpha}_{l}$ ). The reason for the sign difference of $\alpha
_{l}$ in the expression of $\tilde{\alpha}_{l}$ in (\ref{alf1}) and in the
corresponding expression of Ref. \cite{Bell10} is that, in the latter
reference, for the evaluation of the VEV the negative-energy modes have been
used, with the eigenvalues $k_{l}^{(+)}=2\pi (n_{l}+\alpha _{l})/L_{l}$
instead of $k_{l}^{(-)}=2\pi (n_{l}-\alpha _{l})/L_{l}$. This means that, in
fact, the formulas given in \cite{Bell10} are for the periodicity conditions
(\ref{bc}) with $\alpha _{l}$ replaced by $-2\pi \alpha _{l}$ (see also the
comment in \cite{Bell12}). Comparing (\ref{jfM}) with (\ref{jscMink}), we
see that the relation (\ref{jfjsc}) between the fermionic and scalar current
densities holds in Minkowski spacetime for massive fields as well. Hence, in
the supersymmetric models on the background of the Minkowski spacetime with
equal number of scalar and fermionic degrees of freedom the total current
density vanishes. Note that this is not the case for the currents in the
background of dS spacetime.

Now let us consider the fermionic current density in the asymptotic regions
of the ratio $L_{r}/\eta $. For $L_{r}/\eta \ll 1$ the dominant contribution
to (\ref{jrf2}) comes from the region with large values of $z\eta $. By
taking into account that for large $x$ one has $\left[ I_{-1/2-i\alpha
m}(x)+I_{1/2+i\alpha m}(x)\right] K_{1/2+i\alpha m}(x)\approx 1/x$, we see
that to the leading order $\left\langle j^{r}\right\rangle $ coincides with
the corresponding result for a massless fermionic field $\left\langle
j^{r}\right\rangle \approx (\eta /\alpha )^{D+1}\left\langle
j^{r}\right\rangle ^{\mathrm{(M)}}$. In this region we have the relation (%
\ref{jfjsc}) between the scalar and fermionic field currents.

For small values of the ratio $\eta /L_{r}$, the dominant contribution to
the integral in (\ref{jrf2}) comes from the range with small values of $%
z\eta $. By making use of the expansions for the modified Bessel functions
for small values of the arguments, to the leading order we find%
\begin{equation}
\langle j^{r}\rangle \approx -\frac{NB_{f}e^{-(D+1)t/\alpha }}{2^{p/2}\pi
^{(p+3)/2}V_{q}L_{r}^{p}}\cos [2mt+2\alpha m\ln (L_{r}/\alpha )+\phi _{f}].
\label{jflate}
\end{equation}%
Here, $B_{f}$ and $\phi _{f}$ are defined by the relation%
\begin{equation}
B_{f}e^{i\phi _{f}}=2^{i\alpha m}\Gamma \left( 1/2+i\alpha m\right)
\sum_{n=1}^{\infty }\frac{\sin (n\tilde{\alpha}_{r})}{n^{p+1-2i\alpha m}}%
\sum_{\mathbf{n}_{q-1}^{(r)}}f_{p/2+1-i\alpha m}(nL_{r}u_{r}).  \label{Bf}
\end{equation}%
Hence, unlike the scalar case, the damping of the fermionic current density
for large values of the $L_{r}/\eta $ is always oscillatory. Another
difference is that the oscillation frequency does not depend on the
curvature scale of dS spacetime and is completely determined by the mass of
the field quanta.

\begin{figure}[tbph]
\begin{center}
\epsfig{figure=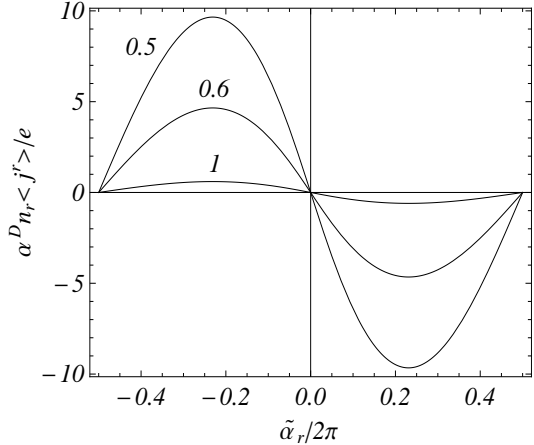,width=7.cm,height=6.cm}
\end{center}
\caption{The current density as a function of the phase in the
quasiperiodicity condition for a fermionic field in the $D=4$ model with a
single compact dimension. The graphs are plotted for $\protect\alpha m=0.25$
and the numbers near the curves are the corresponding values for the ratio $%
L_{r}/\protect\eta $. }
\label{fig3}
\end{figure}

Figure \ref{fig3} presents the quantity $\alpha ^{D}n_{r}\left\langle
j^{r}\right\rangle /e$ for a fermionic field versus the phase in the
quasiperiodicity condition for the $D=4$ model with a single compact
dimension ($p=3$). The graphs are plotted for $\alpha m=0.25$ and the
numbers near the curves are the values for the ratio $L_{r}/\eta $.

\begin{figure}[tbph]
\begin{center}
\begin{tabular}{cc}
\epsfig{figure=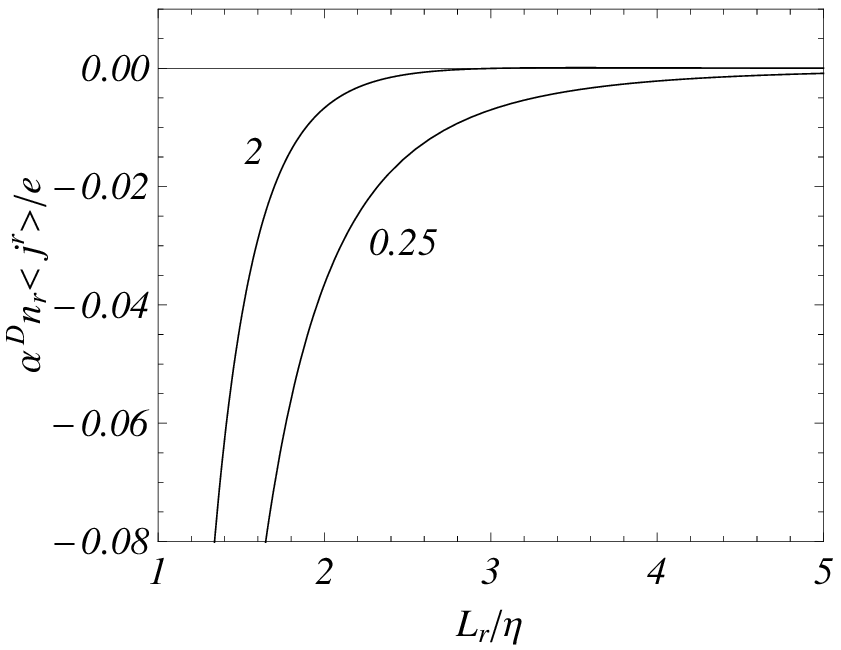,width=7.cm,height=6.cm} & \quad %
\epsfig{figure=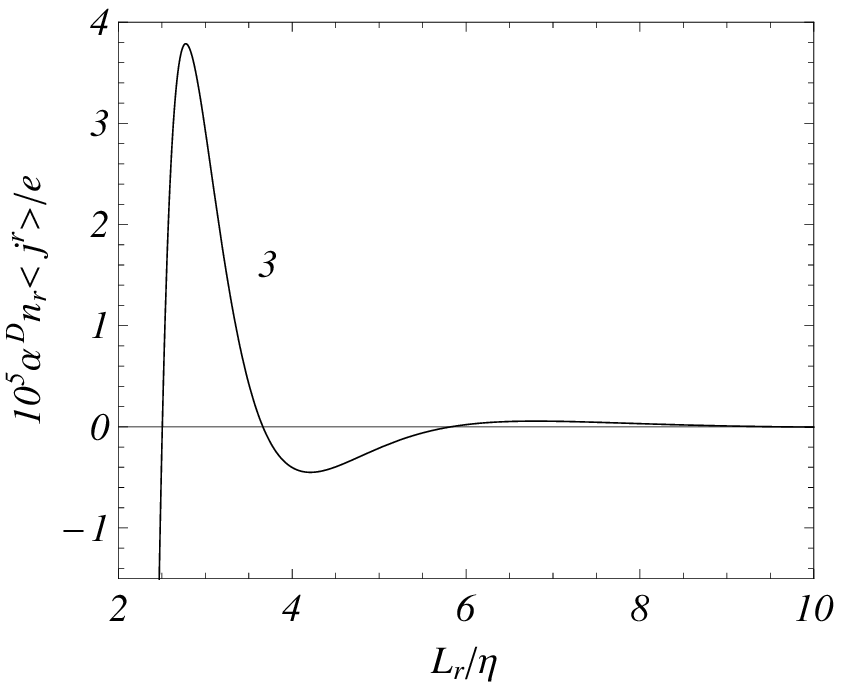,width=7.cm,height=6cm}%
\end{tabular}%
\end{center}
\caption{Fermionic current density as a function of the proper length of the
compact dimension for separate values of the parameter $m \protect\alpha $
(the numbers near the curves). The graphs are plotted for $\tilde{\protect%
\alpha}_{r}=\protect\pi /2$. }
\label{fig4}
\end{figure}

The dependence of the current density on the proper length of the compact
dimensions is displayed in figure \ref{fig4} for the same model $D=4$, $p=3$%
. The graphs are plotted for the phase in the periodicity condition
corresponding to $\tilde{\alpha}_{r}=\pi /2$ and the numbers near the curves
are the values of the parameter $m \alpha $. For the fermionic field the
damping of the current density for large values of $L_{r}/\eta $ is always
oscillatory. In order to display this behavior, on the right panel of figure %
\ref{fig4} we present the current density for the case $m \alpha =3$. As in
the scalar case, the value of the ratio $L_{r}/\eta $ for the the first zero
of the current density decreases with increasing $m \alpha $. For the first
two zeros we has $L_{r}/\eta =3.04, 5.90$ and $L_{r}/\eta =2.51, 3.67$ in
the cases $m \alpha =2$ and $m \alpha =3$, respectively.

\section{Conclusion}

\label{sec:Conc}

Among the most interesting consequences of the compactification of spatial
dimensions is the appearance of nonzero expectation values for physical
observables in the vacuum state of a quantum field. In the present paper we
have considered the effect of the nontrivial topology on the VEVs of the
current density for charged scalar and fermionic fields in the background of
dS spacetime. In both cases we have assumed that the fields are prepared in
the Bunch-Davies vacuum state. Along compact dimensions the quasiperiodicity
conditions (\ref{BC_pq}) and (\ref{bc}) are imposed with general phases. In
addition, we have assumed the presence of a constant classical abelian gauge
filed. This leads to the Aharonov-Bohm-like effect on the current density.
The physical quantities depend on the phases in the periodicity conditions
and on the gauge potential in the form of the combination (\ref{alf1}). The
VEVs of the charge density and of the components of the current density
along compact dimensions vanish.

For the general case of toroidal spatial topology $R^{p}\times (S^{1})^{q}$,
the VEVs for the scalar and fermionic current densities along the $r$th
compact dimension are given by the expressions (\ref{jscr}) and (\ref{jrf2}%
), respectively. These VEVs are even periodic functions of the phases $%
\tilde{\alpha}_{l}$, $l\neq r$, with the period $2\pi $, and they are odd
periodic functions of the phase $\tilde{\alpha}_{r}$ with the same period.
In particular, they exhibit Aharonov-Bohm-type oscillations as functions of
the magnetic flux (with the period of the flux quantum) enclosed by the
compact dimension. The current densities vanish in the special cases of
untwisted and twisted fields, when the gauge field is absent. We have
explicitly checked the transition to the Minkowskian results, previously
discussed in \cite{Beze12,Bell10,Bell12}, in the limit $\alpha \rightarrow
\infty $, for a fixed value of $t$. In this limit, if the phases in the
periodicity conditions and the masses for scalar and fermionic fields are
the same, we have the simple relation (\ref{jfjsc}) between the scalar and
fermioinic current densities. In particular, in supersymmetric models on the
background of the Minkowski spacetime with equal number of scalar and
fermionic degrees of freedom the total current density vanishes. The
background gravitational field modifies the current densities for scalar and
fermionic fields in different ways, and we have no similar cancellation in
the dS spacetime.

If the proper length of the compact dimension is much smaller than the dS
curvature scale, the effects induced by gravity are small and the current
densities are related to the corresponding results for massless fields in
Minkowski spacetime with compact dimensions by the simple formula $%
\left\langle j^{r}\right\rangle \approx (\eta /\alpha )^{D+1}\left\langle
j^{r}\right\rangle ^{\mathrm{(M)}}$. For a fixed value of the ratio $%
L_{r}/\alpha $, this limit corresponds to the early stages of the
cosmological expansion ($t\rightarrow -\infty $), and the current densities
behave as $e^{-(D+1)t/\alpha }$. The effect of gravity on the VEVs is
decisive for proper lengths of the compact dimensions larger than the dS
curvature radius. In this limit, for the case of a scalar field, depending
on the mass, two regimes are realized. For the first one, corresponding to
real values of the parameter $\nu $ defined by (\ref{nu}), the leading term
in the asymptotic expansion of the current density is given by (\ref{jsclate}%
) and the VEVs decay monotonically as $(\eta /L_{r})^{D-2\nu +2}$. For a
fixed value of $L_{r}/\alpha $, this corresponds to late stages of the
cosmological evolution ($t\rightarrow +\infty $), and the current density is
suppressed by the factor $e^{-(D-2\nu +2)t/\alpha }$. In the second regime,
realized for imaginary values of $\nu $, the behavior of the scalar current
density is oscillatory, with the amplitude decaying as $(\eta /L_{r})^{D+2}$%
. The period of oscillations is given by $\pi \alpha /|\nu |$. For the
fermionic field the oscillatory regime is realized only (see (\ref{jflate}))
with the amplitude decaying as $(\eta /L_{r})^{D+1}$. In the fermionic case
the period of oscillations does not depend on the curvature radius and is
given by $2\pi /m$. Note that the decay rates for the amplitude in the cases
of the scalar and fermionic currents are different. At late stages of the
cosmological expansion, the current densities are suppressed by the factors $%
e^{-(D+2)t/\alpha }$ and $e^{-(D+1)t/\alpha }$ for scalar and fermionic
fields, respectively. For the values of the curvature scale $\alpha \gtrsim
m^{-1}$, if the length of the compact dimension is much larger than the
Compton wavelength of the field quanta, $mL_{r}\gg 1$, the decay of the
current densities as functions of $mL_{r}$ is a power-law, for both scalar
and fermionic fields. This is in contrast to the case of the Minkowski bulk,
where the VEVs of the current densities are suppressed exponentially.

The effects we have discussed can be applied to two types of models. For the
first one $D=3$ and the results given here describe how the properties of
the universe are changed by one-loop quantum effects induced by the
compactness of spatial dimensions. Though the observational data constrain
the size of possible compact dimensions to be larger than the horizon scale,
at early stages of the cosmological expansion the lengths of compact
dimensions were small and the effects considered here can be important. The
second class of models, with $p=3$ and $D>3$, correspond to the universe
with Kaluza-Klein-type extra dimensions. In these models, the current
density along compact dimensions could be a source for cosmological magnetic
fields. The gravitational analog of this electromagnetic effect was
discussed in \cite{DarkEn}, where it was shown that the topological Casimir
energy can be considered as a possible origin for the dark energy.

\section*{Acknowledgments}

A.A.S. gratefully acknowledges the hospitality of the INFN, Laboratori
Nazionali di Frascati (Frascati, Italy) where part of this work was done.


\begin{thebibliography}{99}
\bibitem{Birr82} N.D. Birrell and P.C.W. Davies, \textit{Quantum Fields in
Curved Space} (Cambridge University Press, Cambridge, 1982); S.A. Fulling,
\textit{Aspects of Quantum Field Theory in Curved Spacetime} (Cambridge
University Press, Cambridge, 1989); A.A. Grib, S.G. Mamayev, and V.M.
Mostepanenko, \textit{Vacuum Quantum Effects in Strong Fields} (Friedmann
Laboratory Publishing, St. Petersburg, 1994); R.M. Wald, \textit{Quantum
Field Theory in Curved Spacetime and Black Hole Thermodynamics} (Chicago,
University of Chicago Press, 1994); L.E. Parker and D.J. Toms, \textit{%
Quantum Field Theory in Curved Spacetime} (Cambridge University Press,
Cambridge, 2009);

\bibitem{Lind90} A.D. Linde, \textit{Particle Physics and Inflationary
Cosmology} (Harwood Academic Publishers, Chur, Switzerland 1990).

\bibitem{Ries07} A.G. Riess et al., Astrophys. J. \textbf{659}, 98 (2007);
D.N. Spergel et al., Astrophys. J. Suppl. Ser. \textbf{170}, 377 (2007);
U.~Seljak, A. Slosar, and P. McDonald, JCAP \textbf{0610}, 014 (2006); E.
Komatsu et al., arXiv:0803.0547.

\bibitem{Kach03} S. Kachru, R. Kallosh, A. Linde, and S.P. Trivedi, Phys.
Rev. D \textbf{68}, 046005 (2003); E. Silverstein, Phys. Rev. D \textbf{77},
106006 (2008).

\bibitem{Lind04} A. Linde, JCAP \textbf{0410}, 004 (2004).

\bibitem{Zeld84} Y.B. Zeldovich and A.A. Starobinsky, Sov. Astron. Lett.
\textbf{10}, 135 (1984).

\bibitem{Gonc85} Yu.P. Goncharov and A.A. Bytsenko, Phys. Lett. B \textbf{160%
}, 385 (1985); Yu.P. Goncharov and A.A. Bytsenko, Nucl. Phys. B \textbf{271}%
, 726 (1986); Yu.P. Goncharov and A.A. Bytsenko, Class. Quant. Grav. \textbf{%
4}, 555 (1987).

\bibitem{McIn04} B. McInnes, Nucl. Phys. B \textbf{692}, 270 (2004); B.
McInnes, Nucl. Phys. B \textbf{709}, 213 (2005); B. McInnes, Nucl. Phys. B
\textbf{748}, 309 (2006).

\bibitem{Most97} V.M. Mostepanenko and N.N. Trunov, \textit{The Casimir
Effect and Its Applications} (Clarendon, Oxford, 1997); K.A. Milton, \textit{%
The Casimir Effect: Physical Manifestation of Zero-Point Energy} (World
Scientific, Singapore, 2002); M. Bordag, G.L. Klimchitskaya, U. Mohideen,
and V.M. Mostepanenko, \textit{Advances in the Casimir Effect} (Oxford
University Press, Oxford, 2009); \textit{Lecture Notes in Physics: Casimir
Physics,} Vol. 834, edited by D. Dalvit, P. Milonni, D. Roberts, and F. da
Rosa (Springer, Berlin, 2011).

\bibitem{Duff86} M.J. Duff, B.E.W. Nilsson, and C.N. Pope, Phys. Rep.
\textbf{130}, 1 (1986); R. Camporesi, Phys. Rep. \textbf{196}, 1 (1990);
A.A. Bytsenko, G. Cognola, L. Vanzo, and S. Zerbini, Phys. Rep. \textbf{266}%
, 1 (1996); A.A. Bytsenko, G. Cognola, E. Elizalde, V. Moretti, and S.
Zerbini, \textit{Analytic Aspects of Quantum Fields} (World Scientific,
Singapore, 2003); E. Elizalde, \textit{Ten Physical Applications of Spectral
Zeta Functions} (Springer Verlag, 2012).

\bibitem{DarkEn} E. Elizalde, Phys. Lett. B \textbf{516}, 143 (2001); C.L.
Gardner, Phys. Lett. B \textbf{524}, 21 (2002); K.A. Milton, Grav. Cosmol.
\textbf{9}, 66 (2003); A.A. Saharian, Phys. Rev. D \textbf{70}, 064026
(2004); E. Elizalde, J. Phys. A \textbf{39}, 6299 (2006); A.A. Saharian,
Phys. Rev. D \textbf{74}, 124009 (2006); B. Green and J. Levin, JHEP \textbf{%
0711}, 096 (2007); P. Burikham, A. Chatrabhuti, P. Patcharamaneepakorn, and
K. Pimsamarn, JHEP \textbf{0807}, 013 (2008); P. Chen, Nucl. Phys. B (Proc.
Suppl.) \textbf{173}, s8 (2009).

\bibitem{Saha07} A.A. Saharian and M.R. Setare, Phys. Lett. B \textbf{659},
367 (2008).

\bibitem{Bell08} S. Bellucci and A.A. Saharian, Phys. Rev. D \textbf{77},
124010 (2008).

\bibitem{Saha08} A.A. Saharian, Class. Quantum Grav. \textbf{25}, 165012
(2008).

\bibitem{Beze09} E.R. Bezerra de Mello and A.A. Saharian, J. High Energy
Phys. \textbf{04}, 046 (2009).

\bibitem{Witt01} E. Witten, arXiv: hep-th/0106109.

\bibitem{Klem04} D. Klemm and L. Vanzo, JCAP \textbf{0411}, 006 (2004).

\bibitem{Ande03} L. Andersson and G.J. Galloway, Adv. Theor. Math. Phys.
\textbf{6}, 307 (2003); G.J. Galloway, arXiv:gr-qc/0407100.

\bibitem{Degu13} H. Degueldre, R. P. Woodard, arXiv:1303.3042.

\bibitem{Seib99} N. Seiberg and E. Witten, J. High Energy Phys. \textbf{04},
017 (1999).

\bibitem{Hoso83} Y. Hosotani, Phys. Lett. B \textbf{126}, 309 (1983); A.
Higuchi and L. Parker, Phys. Rev. D \textbf{37}, 2853 (1988); Y. Hosotani,
Ann. Phys. \textbf{190}, 233 (1989); A. Actor, Class. Quantum Grav. \textbf{7%
}, 663 (1990); K. Kirsten, J. Phys. A \textbf{26}, 2421 (1993).

\bibitem{Alle85} B. Allen, Phys. Rev. D \textbf{32}, 3136 (1985); B. Allen
and A. Folacci, Phys. Rev. D \textbf{35}, 3771 (1987).

\bibitem{Bunc78} T.S. Bunch and P.C.W. Davies, Proc. R. Soc. London, Ser. A
\textbf{360}, 117 (1978).

\bibitem{Beze08} E.R. Bezerra de Mello and A.A. Saharian, Phys. Rev. D
\textbf{78}, 045021 (2008).

\bibitem{Bell09} S. Bellucci and A.A. Saharian, Phys. Rev. D \textbf{79},
085019 (2009).

\bibitem{SahaBook} A.A. Saharian, \textit{The Generalized Abel-Plana Formula
with Applications to Bessel Functions and Casimir Effect} (Yerevan State
University Publishing House, Yerevan, 2008); Report No. ICTP/2007/082;
arXiv:0708.1187.

\bibitem{Beze12} E.R. Bezerra de Mello and A.A. Saharian, Phys. Rev. D, in
press, arXiv:1211.5174.

\bibitem{Duns90} T.M. Dunster, SIAM J. Math. Anal. \textbf{21}, 995 (1990);
K.A. Milton, J. Wagner, and K. Kirsten, Phys. Rev. D \textbf{80}, 125028
(2009).

\bibitem{Prud86} A.P. Prudnikov, Yu.A. Brychkov, and O.I. Marichev, \textit{%
Integrals and Series} (Gordon and Breach, New York, 1986), Vol. 2.

\bibitem{Bell10} S. Bellucci, A.A. Saharian, and V.M. Bardeghyan, Phys. Rev. D \textbf{82},
065011 (2010).

\bibitem{Bell12} S. Bellucci and A.A. Saharian, Phys. Rev. D \textbf{87},
025005 (2013).
\end{thebibliography}
\end{document}